\documentclass[12pt]{article}

\usepackage{epsf,amssymb}

\newcommand{\zr}[1]{\mbox{\hspace*{#1em}}}

\newcommand{\NN}{\mathbb{N}}
\newcommand{\ZZ}{\mathbb{Z}}
\newcommand{\QQ}{\mathbb{Q}}

\begin{document}
 \bibliographystyle{unsrt}

 \parindent0pt
 
 \begin{center}
 {\Large {\bf Some remarks on the visible points of a lattice }}
 \end{center}
 \vspace{3mm}
 
 \begin{center}
 {\bf {\sc M.~Baake$^{1),2)}$, U.~Grimm$^{2),3)}$, and D.~H.~Warrington$^{4)}$}}
 \end{center}
\vspace{5mm}

 {\footnotesize
 \hspace*{6em} 1) Institut f\"ur Theoretische Physik, Universit\"at T\"ubingen, \\
 \hspace*{6em} \hspace*{1em} Auf der Morgenstelle 14, 72076 T\"ubingen, Germany
 
 \hspace*{6em} 2) Department of Mathematics, The University of Melbourne, \\
 \hspace*{6em} \hspace*{1em} Parkville, VIC 3052, Australia

 \hspace*{6em} 3) Instituut voor Theoretische Fysica, Universiteit van Amsterdam, \\
 \hspace*{6em} \hspace*{1em} Valckenierstraat 65, 1018 XE Amsterdam, The Netherlands

 \hspace*{6em} 4) School of Materials, The University of Sheffield, \\
 \hspace*{6em} \hspace*{1em} Mappin Street, Sheffield S1 3JD, Great Britain

  $\mbox{ }$}
 
 



 \vspace{5mm}
 
\begin{abstract}
We comment on the set of visible points of a lattice and its Fourier
transform, thus continuing and generalizing
previous work by Schroeder \cite{Sch1}
and Mosseri \cite{Mossy}.
A closed formula in terms of Dirichlet series is obtained for the
Bragg part of the Fourier transform. We compare this calculation with
the outcome of an optical 
Fourier transform of the visible points of the 2D square lattice.
\end{abstract}

\parindent15pt
\vspace{5mm}

Recently, Mosseri has given a nice and elegant description of the set 
of visible points 
of a lattice (i.e., those points except the origin that connect to
the origin via a straight line without hitting any other lattice point
in between) 
and some of its properties \cite{Mossy}, continuing previous work by
Schroeder \cite{Sch1,Sch2}. Both authors are most interested in the
Fourier transform of this set.
Unfortunately, the results of these two attempts are contradictory.
The algebraic approach of \cite{Mossy} is essentially correct,
but contains a couple of mistakes and unnecessary restrictions.
In what follows, we add several 
remarks to straighten this out and then compare the 
2D case with an optical experiment.

Let $\Lambda = \ZZ b_1 \oplus \cdots \oplus \ZZ b_n$ be a lattice in $n$D space with
linearly independent basis vectors $b_1, \ldots, b_n$. The set $F_{\Lambda}$ of visible
points \cite{Hardy,Sch2,Mossy} can be characterized as
\begin{equation} \label{1.0}
   F_{\Lambda} = \{m_1 b_1 + \cdots + m_n b_n \; | \; gcd(m_1, \ldots , m_n) = 1 \},
\end{equation}
where $gcd$ denotes the greatest common divisor.
This set does {\em not} include the origin with respect to which it is defined. We follow
\cite{Mossy} for the notation as far as possible.

The set $F_\Lambda$ is non-periodic and is left invariant by the group 
of lattice automorphisms, $Aut(\Lambda)$,
which is isomorphic to $Gl(n,\ZZ)$, the group of integer 
$n \! \times \! n$ matrices with
determinant $\pm 1$. This is seen from $M \in Aut(\Lambda)$ transforming fundamental cells
of the lattice to other fundamental cells and hence visible points to visible points
\cite{Hardy,Siegel}. As a consequence, the set of visible points admits precisely the same
point symmetry as the lattice $\Lambda$ itself, the transformations are not restricted
to pure rotations. This can clearly be seen from Fig.\ 1 which shows the case of the
square lattice.

How frequent are visible points? If $p$ denotes the probability of a lattice point 
to be visible (defined through a volume limit which exists, compare
chapter 3.8 of \cite{Apostol2}), 
we also have ($\ell \in \NN=\{1,2,3,\ldots\}$) the probabilities
\begin{equation} \label{2.0}
    P(x \in \ell \! \cdot \!  F_{\Lambda}) = \frac{p}{\ell^n} \,\, ,
\end{equation}
because $\ell \!  \cdot \!  F_{\Lambda} = 
\{m_1 b_1 + \cdots + m_n b_n \; | \; gcd(m_1, \ldots , m_n) = \ell \}$.
On the other hand, we have
\begin{equation} \label{3.0}
   \Lambda = \{0\} \, \cup \, \bigcup_{\ell=1}^{\infty} \ell \!  \cdot \!  F_{\Lambda} \,\, ,
\end{equation}
which is the correct version of eq.~(1) in \cite{Mossy}. 
Note that (\ref{3.0}) is the union of pairwise disjoint
sets\footnote{This property would be destroyed by the (unconventional)
addition of the origin to the set $F_{\Lambda}$.} which will be
essential in what follows.

The single point $0$, however,
does not matter in the calculation of the probabilities wherefore we get
\begin{equation} \label{4.0}
    1 = \sum_{\ell=1}^{\infty} P(x \in \ell \! \cdot \! F_{\Lambda}) 
      = p \cdot \sum_{\ell=1}^{\infty} \frac{1}{\ell^n} \;.
\end{equation}
We can thus write $p$ by means of Riemann's $\zeta$-function as
\begin{equation} \label{5.0}
   p = \frac{1}{\zeta(n)} = \prod_{p \in {\cal P}} (1 - \frac{1}{p^n}) \,\, ,
\end{equation}
where ${\cal P}$ denotes the set of primes \cite{Hardy,Titmarsh}.
Obviously, $p=0$ for $n=1$ (precisely two points are visible here), and $p$ rapidly
approaches 1 with increasing $n$, compare Fig.\ 2. 
For even $n \geq 2$, the $\zeta$-function is known
to be transcendental \cite{Zeta,Apostol} (and thus irrational) through
$\zeta(2m) = \frac{(2\pi)^{2m}}{2 (2m)!} |B_{2m}|$, where $B_{2m}$ are the Bernoulli
numbers, compare \cite{Zeta}.
Also, $\zeta(3)$ is irrational \cite{Apery}, while this question is
open for odd $n > 3$. Note that the nonperiodicity of $F_{\Lambda}$ follows
from the irrationality of $p$, but not vice versa. 

A different approach uses the 3rd M\"obius inversion formula \cite{Sch2,Hardy} where one starts from
the characteristic function $h_{\Gamma}$ of a set $\Gamma$, which is one for points of the set and
zero otherwise, and rewrites Eq.~(\ref{3.0}) as
\begin{equation} \label{6.0}
   h_{\Lambda} = h_{\{0\}} + \sum_{\ell=1}^{\infty} h_{\ell \cdot F_{\Lambda}}.
\end{equation}
Convergence is no problem because, for any given point, the right hand
side is a finite sum.
Now, using the inversion formula for such series \cite{Hardy} one obtains
\begin{equation} \label{7.0}
   h_{F_{\Lambda}} = \sum_{\ell=1}^{\infty} \mu(\ell) h_{\ell \cdot \Lambda  \, \backslash \, \{0\}}
\end{equation}
with the M\"obius function $\mu(\ell)$, cf. \cite{Hardy}. Note the differences 
to eqs.~(1-3)
in \cite{Mossy} which result from the little mistake in the description 
of the lattice $\Lambda$.
It is important though because $|\sum_{\ell=1}^{N} \mu(\ell)|$ does 
not converge:
it diverges faster than $N^{1/2}$ \cite{Sch2,Odlyzko}. 

This indicates that the calculation of the Fourier transform requires some care: it
is not clear that the set of visible points is strictly almost periodic
in the sense that the Fourier transform need not be a sum over
$\delta$-distributions (or Bragg peaks) only.
In fact, for $n=1$, this diffractive part vanishes due to the 
existence of only 2 visible points
and the Fourier transform is $2 \cos(kx)$ and thus continuous.
This limit must nevertheless be covered by the correct treatment.
In \cite{Mossy}, this is hidden by an infinite normalization factor
-- the formula would give vanishing Fourier transform for $n=1$. 

Let us now consider the so-called structure factor in more detail.
The structure factor $S_{\Gamma}$ of a discrete set $\Gamma$ is the Fourier 
transform of $\delta$-scatterers of equal strength on all points of $\Gamma$, i.e.,
\begin{equation} \label{8.0}
   S_{\Gamma}(k) = \int \sum_{x\in\Gamma} \delta(x'-x) e^{-ikx'} dx' = \sum_{x\in\Gamma} e^{-ikx}.
\end{equation}
If $\Gamma$ has inversion symmetry (as $\Lambda$ and $F_{\Lambda}$ 
do), we can also write $S_{\Gamma} (k) = \sum_{x\in\Gamma} e^{ikx} = \sum_{x\in\Gamma} \cos(kx)$.
Now, with Eq.~(\ref{7.0}), we find
\begin{equation} \label{9.0}
   S_{F_{\Lambda}}(k) = \sum_{x \in \Lambda} \,\,\,\sum_{\ell=1}^{\infty} \mu(\ell) 
   h_{\ell \cdot \Lambda \backslash \{0\}} (x) \cdot e^{ikx} \, ,
\end{equation}
to be understood in the distribution sense. Nevertheless, due to lack of convergence,
an interchange of the sums is a subtle business.
Though a complete treatment is desirable here, it is far beyond the
scope of this comment and therefore left to a forthcoming publication
\cite{new}.
But in the spirit of \cite{Mossy}, one can indeed extract the point-like
contributions, henceforth called Bragg peaks (though we are talking about
{\em amplitudes} rather than intensities here) because other
contributions are arbitrarily small in comparison to the $\delta$-peaks
in the thermodynamic limit.

Before we continue, let us remark that the structure factor
$S_{F_{\Lambda}}(k)$ is {\em periodic} in $k$ w.r.t.\ the reciprocal
lattice $\Lambda^*$: we can alternatively write
\begin{equation} \label{7.1}
  S_{F_{\Lambda}}(k) = \int \sum_{x \in \Lambda} \delta(x'-x) \cdot
    h_{F_{\Lambda}}(x') e^{-ikx'} dx'
\end{equation}
with the characteristic function $h_{F_{\Lambda}}$ defined above.
But then, $S_{F_{\Lambda}}(k)$ is the convolution of $S_{\Lambda}(k)$
-- which is periodic ! -- with the Fourier transform of the
characteristic function $h_{F_{\Lambda}}$. Consequently, 
$S_{F_{\Lambda}}(k)$ itself is periodic.
Furthermore, since the set $F_{\Lambda}$ of visible points is invariant under
$Aut(\Lambda)$, the corresponding property applies to its Fourier
transform, which is thus invariant under $Aut(2 \pi \Lambda^*)$ -- the
latter again being isomorphic with the group $Gl(n,\ZZ)$.
These two arguments apply to the full structure factor (and also to
$n=1$), not only to its Bragg peaks.

To describe the latter, we now consider the reciprocal lattice 
$2 \pi \Lambda^*$ with $\Lambda^*=\{y \; | \; x \! \cdot \! y \in
\ZZ \hspace{3mm}\forall x \in \Lambda \} = \ZZ b_1^* + \cdots + 
\ZZ b_n^*$, $b_i^* \! \cdot \! b_j = \delta_{ij}$,
a Bragg peak sits on every 
$k=2\pi(r_1 b_1^* + \cdots + r_n b_n^*)$ with {\em rational} coefficients, i.e., $r_i \in \QQ$.
If we write $r_i=p_i/q_i$ with coprime integers $p_i,q_i$ (where we take
$q_i > 0$ for convenience), the peaks have
strength $H=H(q_1,\ldots,q_n)$,
\begin{equation} \label{11.0}
   H(q_1,\ldots,q_n) = \sum_{m=1}^{\infty} \frac{\mu(ma)}{(ma)^n} \;\; , \; 
   a=lcm(q_1,\ldots,q_n),
\end{equation}
relative to the Bragg peaks of the lattice $\Lambda$.
Here, $lcm$ denotes the least common multiple. 
In particular, $q_1=\ldots=q_n=1$ gives
\begin{equation} \label{12.0}
   H(1,...,1) = \sum_{m=1}^{\infty} 
  \frac{\mu(m)}{m^n} = \frac{1}{\zeta(n)} \, ,
\end{equation}
which is again the frequency of visible points,
while $H$ is zero unless {\em all} arguments are square free, 
because the M\"obius function vanishes for numbers that 
are divisible by a square.
This explains the ``dark'' lines in the diagram of intensities 
(i.e., lines without spots -- hence ``white'' in Fig.\ 3 while really
dark in Fig.\ 4)  which are given 
-- in kinematic approximation -- by the absolute squares of the amplitudes
(\ref{11.0}). 

Eq.~(\ref{11.0}) gives a closed formula for the amplitudes of the 
Bragg peaks in form of an infinite sum,
but can still be simplified considerably. If we consider
\begin{equation} \label{13.0}
   f(n,a) = \sum_{\ell=1}^{\infty} \frac{\mu(\ell a)}{(\ell a)^n}
\end{equation}
for $\ell,a \in \NN$, we can use standard techniques from the treatment of
Dirichlet series, compare \cite{Titmarsh}, to evaluate the sum.
If $a=p_1 \cdot \ldots \cdot p_r$ is the product of pairwise distinct
primes, one can easily show by induction that
\begin{equation} \label{14.0}
   f(n,p_1 \cdot \ldots \cdot p_r) =
    \frac{f(n,1)}{\prod_{j=1}^{r} (1 - p_j^n) } \,\, ,
\end{equation}
and hence, with the multiplicativity of the M\"obius function and Eq.~(\ref{13.0}),
\begin{equation} \label{15.0}
   f(n,a) = \frac{\mu(a)}{a^n} \cdot
            \prod_{p\not\zr{0.17}\mid\zr{0.17} a, \, p \in {\cal P}} 
            \,\, (1 - \frac{1}{p^n}) =
            \frac{1}{\zeta(n)} \cdot \frac{\mu(a)}
           { \prod_{p \mid a, \, p \in {\cal P}} \,\, (p^n - 1)} \,\, .
\end{equation}
Inserting this into (\ref{11.0}) finally gives, with $a=lcm(q_1,\ldots,q_n)$,
the amplitude function
\begin{equation} \label{16.0}
   H(q_1,\ldots,q_n) = H(1,\ldots,1) \cdot \frac{\mu(a)}
           { \prod_{p \mid a, \, p \in {\cal P}} \,\, (p^n - 1)} \,\, .
\end{equation}
This way, the calculation of the amplitudes {\em relative}
to the central amplitude is reduced to the evaluation of a 
finite product, compare Fig.\ 3 for the 2D example of the square lattice.
There, a Bragg peak is represented by a dot with radius proportional
to the amplitude (i.e., the area is proportional to the intensity).
   
The Fourier transform is periodic, although the set itself obviously is not
-- it is not even quasiperiodic because the Fourier module (i.e., the
$\ZZ$-module generated by the positions of the $\delta$-peaks) is only
countably but not finitely generated.
The key structure is thus contained in one fundamental domain of the reciprocal lattice.
This is indeed clearly seen in Figs.\ 3 and 4. The latter was obtained
by optical Fourier transform of a finite portion of 
$F_{\ZZ^2_{}}^{}$ on an optical bench.
Here, we used some 9000 points of radius 15 $\%$ of the lattice constant.
They were prepared as a negative slide made from $24 \! \times \! 36$
mm$^2$ documentary film. The light was extracted from a laser beam,
widened by a high quality lens to a parallel beam covering the entire
slide.
Although the image is not perfect (one should not expect rapid convergence
in view of the remarks after (\ref{9.0}))
and still shows some standard optical errors, it is nevertheless amazing how 
clear the structure with the subtle invariance properties is recovered.
It is correctly described by the approach via the M\"obius transform 
while previous numerical 
attempts \cite{Sch1} seem to fail for reasons we were not able to unravel.

Of course, one can now deal with generalizations like the set of points 
visible to two or more observers which resembles the above situation in 
many respects, although the relative position of the observers does play 
a significant role. Unfortunately, we could not find similarly nice closed 
expressions for the Fourier transform wherefore we drop further details here.

\vspace{5mm}
\parindent 0pt
{\bf Acknowledgements}
\vspace{2mm}
\parindent 15pt

We thank P.~A.~B.\ Pleasants, M.~Schlottmann and C.\ Sire for interesting 
discussions on the subject and J.\ Bachteler and
D.\ Joseph for critically reading the manuscript.
Financial support from Deutsche Forschungsgemeinschaft
is gratefully acknowledged.

\clearpage

\begin{figure}
\centerline{\epsfbox{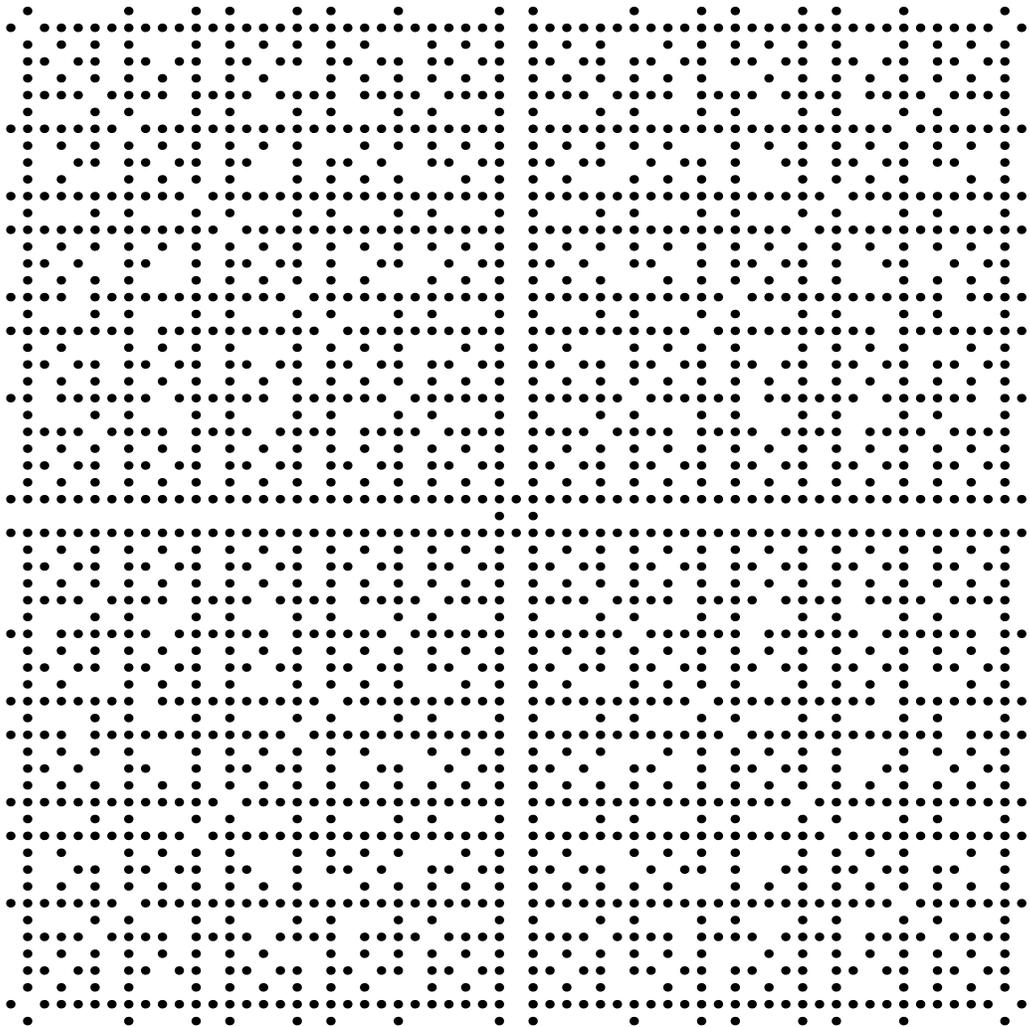}}
\caption{Some visible points of the 2D square lattice.}
\end{figure}\clearpage

\begin{figure}
\centerline{\epsfbox{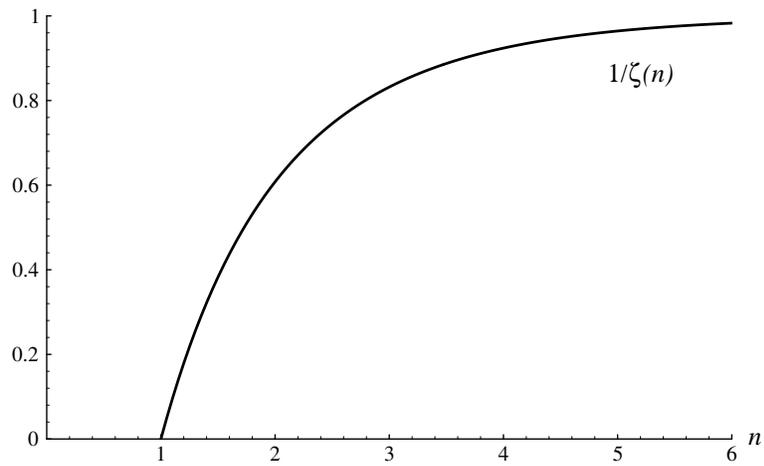}}
\caption{Relative frequency of visible points 
     in a lattice as a function of the dimension.}
\end{figure}\clearpage

\begin{figure}
\centerline{\epsfbox{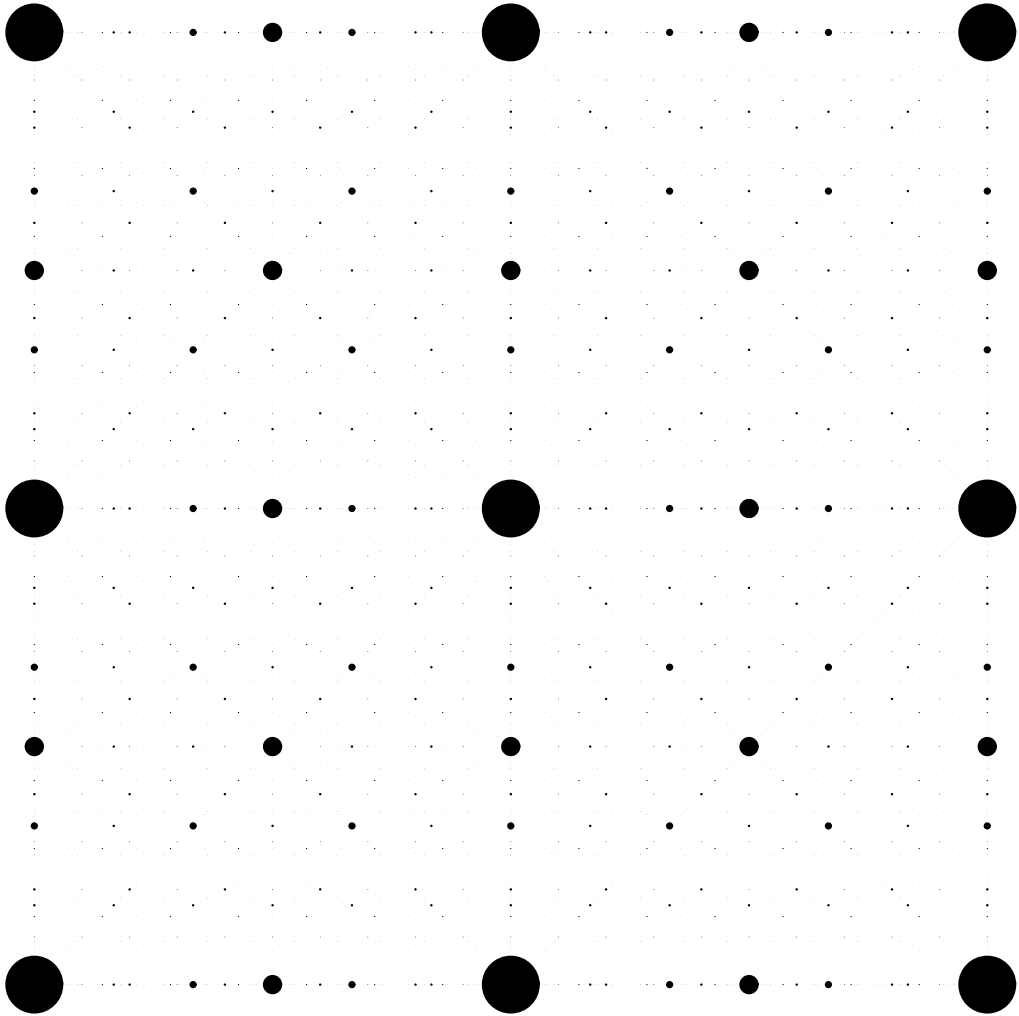}}
\caption{Bragg part of the Fourier transform of the square 
lattice according to (\protect\ref{16.0}). 
Here, the radius of the dots representing
the Bragg peaks are proportional to the amplitude.}
\end{figure}\clearpage

\begin{figure}
\caption{Optical Fourier transform of some 9000 visible points
       of the square lattice $\ZZ^2$, see text for details.}
\end{figure}

\end{document}